\documentclass{aastex}
\usepackage{mathrsfs}
\usepackage{amsfonts}
\usepackage{spr-astr-addons}
\usepackage{url}

\RequirePackage{color}

\newcommand{\emaila}{747352829@qq.com}

\begin{document}

\title{The jets-accretion relation, mass-luminosity relation in Fermi blazars}
\slugcomment{Not to appear in Nonlearned J., 45.}
\shorttitle{Short article title}
\shortauthors{Autors et al.}

\author{Xiaoling Yu} \and \author{Xiong Zhang$^{\dag}$} \and \author{Haojing Zhang} \author{Dingrong Xiong}\and \author{Bijun Li}\and \author{Yongjuan Cha} \and \author{Yongyun Chen} \and \author{Xia Huang} \and \author{Yuwei Wang}\affil{Department of Physics, Yunnan Normal University, 650500,
\\Kunming, China\\$^{\dag}$e-mail:ynzx@yeah.net}
\email{\emaila}

\begin{abstract}
A sample of 111 Fermi blazars each with a well-established radio core luminosity, broad-line luminosity, bolometric luminosity and black hole mass has been compiled from the literatures.We present a significant correlation between radio core and broad-line emission luminosities that supports a close link between accretion processes and relativistic jets. Analysis reveals a relationship of $\rm{LogL_{BLR}\sim(0.81\pm0.06)LogL_{R}^{C}}$ which is consistant with theoretical predicted coefficient and supports that blazar jets are powered by energy extraction from a rapidly spinning Kerr black hole through the magnetic field provided by the accretion disk. Through studying the correlation between the intrinsic bolometric luminosity and the black hole mass, we find a relationship of $\rm{{Log}\frac{L_{in}}{L_{\odot}}=(0.95\pm0.26){Log}\frac{M}{M_{\odot}}+(3.53\pm2.24)}$ which supports mass-luminosity relation for Fermi blazars derived in this work is a powerlaw relation similar to that for main-sequence stars. Finally, EVOLUTIONARY SEQUENCE OF BLAZARS is discussed.
\end{abstract}



\section{INTRODUCTION}
Blazars are the most violent type of active galactic nuclei(AGNs), showing some extreme observational properties, such as high and variable luminosity, superluminal motions in the radio components, high $\gamma$-ray emission, no emission lines or emission lines that are very strong, etc.(e.g. Aller et al. 1992, 2003; Andruchow et al. 2005; Cellone et al. 2007; Fan et al. 2010; Zheng et al. 2008). The extremely observational properties are explained as being due to a beaming effect. The emissions from the jet are strongly boosted in the observes frame. Generally, blazars are divided into two subclasses, which contain BL Lacertae objects(BL Lacs) and Flat Spectrum Radio Quasars(FSRQs). FSRQs have strong emission lines, while BL Lac objects have only very weak or nonexistent emission lines. The typcally division between FSRQs and BL Lacs is mainly based on the Equivalent Width(EW) of the emission lines. Objects with rest frame $\rm{EW}>5~\rm{\AA}$ are classified as FSRQs (see e.g. Urry \& Padovani 1995). The BL Lacs are divided into subclasses as follows: X-ray selected BL Lac objects(XBLs) and radio selected BL Lacs(RBLs) from surveys. Based on the spectral energy distribution(SED), the BL Lacs also can be divided into three subclasses, namely, highly peaked BL Lac object(HBLs), intermediate objects(IBLs) and low frequency peaked BL Lac objects(LBLs)(Padovani \& Giommi 1995). For the BL Lac objects and FSRQs, Ghisellini et al. (2011) proposed a physical distinction. Based on the luminosity of the broad line region measured in Eddington units, and the dividing line is of the order of $\rm{{L_{BLR}}/{L_{Edd}}\sim5\times 10^{-4}}$. The result also was confirmed by Sbarrato et al. (2012).

In astrophysics, the relation between highly relativistic jets and accretion processes in active galactic nuclei (AGN) is one of the most fundamental open problems. In current theoretical models of the formation of the jet, power is generated through accretion and then extracted from the disk /black hole (BH) rotational energy and converted into the kinetic power of the jet (Blandford \& Znajek 1977; Blandford \& Payne 1982). Both of these two cases, the magnetic field must play an important role in channeling power from the BH or the disk into the jet. Both of these scenarios it should be sustained by matter accretion onto the BH, leading to the relation between the accretion power and jet power (Maraschi \& Tavecchio 2003). The jet-disk connection has been extensively explored by many authors and in different ways (e.g., Rawlings \& Saunders 1991; Falcke \& Biermann 1995; Serjeant et al. 1998; Cao\& Jiang 1999; Wang et al. 2004; Liu et al. 2006; Xie et al. 2007; Gu et al. 2009; Ghisellini et al. 2009, 2010, 2011; Sbarrato et al. 2012). Serjeant et al. (1998) found a significant correlation between radio and optical emission for a sample of steep-spectrum quasars. The result presents direct evidence for a close link between accretion onto BHs and the fueling of relativistic jets. Cao \& Jiang (1999) found a significant correlation between radio and broad-line emission for a sample of radio-loud quasars. Czerny et al. (2004) analyzed the consequences of the hypothesis that the formation of broad-line regions (BLRs) is intrinsically connected to the existence of cold accretion disks.

Because of blazars showing some extreme observational properties, so we adopt blazars as the laboratories to study the physics of relativistic jets and accretion disks. We also study the mass-luminosity relation, accretion rate-luminosity relation and evolutionary sequence of blazars. The results may be very interesting. Urry \& Padovani (1995) found their emission (from radio to gamma rays) is dominated by the beamed nonthermal continuum produced in the jet. The luminosity of the broad-line emission can be taken as an indicator of the accretion power of the objects (Celotti et al. 1997), and also known that the radio core luminosity is an indicator of jet power (Blandford \& Konigl 1979). The connection between the subclasses of blazars has been studied by many authors (e.g., Sambruna et al. 1996; Ghiselliniet al. 1998; Xie et al. 2001). But, it is still uncertain whether the different subclasses of blazars are connected through an evolutionary sequence, or whether they are in limited periods of AGN activity on parallel evolutionary paths (B$\ddot{o}$ttcher \& Dermer 2002). Vagnetti, Cavaliere, \& Giallongo (1991) were based on the radio-optical analysis of the evolution which contain a very small number of BL Lacs and suggested that there might be no need for two separate unified schemes for the FSRQs and BL Lac objects. However, Padovani (1992) studied the relation between BL Lacs and FSRQs. He gave the result that an evolutionary connection between the two subclasses is not supported by the intrinsic properties of the two subclasses, that is, the extended radio emissions and line luminosities are significantly different. Ghisellini et al. (1998) studied these differences used a large sample of blazar and found an increase of peak frequency in a sequence of FSRQs and BL Lac objects. But D'Elia \& Cavaliere (2001) proposed another view about these problems. They argued that BL Lac objects are mainly powered by the rotational energy extraction of central supermassive black holes (SMBHs) via the BZ mechanism, while FSRQs require a dominant contribution from the accretion of the central SMBHs.

The Large Area Telescope on board the Fermi $\gamma$-ray Space Telescope has been scanning the entire $\gamma$-ray sky approximately once every three hours since July of 2008. We have entered in a new era of blazars research (Abdo et al. 2009, 2010; Ackermann et al. 2011). Up to now, the second catalog of AGNs detected by LAT has been released. The second LAT AGN catalog (2LAC) includes 1017 $\gamma$-ray sources located at high galactic latitudes ($|$b$|$ $>$ $10^\circ$) that are detected with a test statistic (TS) greater than 25 and associated statistically with AGNs (Ackermann et al. 2011). We define a clean sample which includes 886 AGNs, which have 395 BL Lac objects, 310 FSRQs, 157 candidate blazars which unknown type, 8 misaligned AGNs, 4 narrowline Seyfert 1 (NLS1s), 10 AGNs of other types, and 2 starburst galaxies (Ackermann et al. 2011). The Fermi LAT¡¯s has higher sensitivity than EGRET and nearly uniform sky coverage. So it becomes a powerful tool for investigating the properties of large populations.

In this paper, we study the relation between luminosity of the broad-line emission ($\rm{L_{BLR}}$) and the radio core luminosity ($\rm{L_{R}^{C}}$) at 5 GHz of Fermi blazars. We also study the mass-luminosity relation, the accretion rate-luminosity relation for Fermi blazars and the nature of the FSRQs-BL Lac objects relationship. The structure of this paper is as follows: in section 2 we present the sample. In section 3 we give the results. Our discussions are presented in section 4. We provide conclusion in section 5. In this paper, we adopt the cosmological parameters $\rm{H_{\rm{0}}=70Km s^{-1} Mpc^{-1}}$, $\rm{\Omega_{\rm{m}}=0.3}$, $\rm{\Omega_{\rm{\Lambda}}=0.7}$.

\section{THE SAMPLES}
 We collected the largest samples of Fermi blazars and got broad line luminosity, black hole mass and cross-correlated these samples with clean blazars detected by Fermi LAT. We cross-correlated them with the following samples which get the core flux at 5 GHz. Firstly we got the broad-line data and black hole from Xiong et al. (2014) and they collected the largest samples of Fermi blazars with broad-line data. Secondly, we considered the following samples of blazars to get core flux of 5 GHz: Dodson et al. (2008), Laurent-Muehleisen et al. (1997), Gu et al. (2009), Giroletti et al. (2004), Jorstad et al. (2004), Wajima et al. (2006). At last, we cross-correlated these samples with the samples of Nemmen et al. (2012) and Finke et al. (2013) to get the beaming factor and the synchrotron peak luminosity respectively. In total, we have a sample containing 111 clean Fermi blazars which have 73 FSRQs and 38 BL Lacs.

\subsection{The broad-line luminosity}
Celotti et al. (1997) estimate the total luminosity emitted in broad lines, $\rm{L_{BLR}}$. They considered fluxes for the following lines: Ly$\alpha$, CIV, MgII, H$\gamma$, H$\beta$ and H$\alpha$, which are amongst the major contributors to the total $\rm{L_{BLR}}$(making up in fact about 60 percent of it) and are well identified in quasar spectra. They gave a total observed luminosities in a certain number of broad lines $\rm{\sum_{i}L_{i,obs}}$, the total $\rm{L_{BLR}}$ can be calculated as $\rm{L_{BLR}=\sum_{i}L_{i,obs}\frac{<L^{*}_{BLR}>}{\sum_{i}L^{*}_{i,est}}}$. Where $\rm{\sum_{i}L^{*}_{i,est}}$ was the sum of the luminosities from the same lines, the total $\rm{<L^{*}_{BLR}>}=555.77$. But not all emission lines we can be obtained for the sources. So some authors used one particular broad line multiplied by a scaling factor to obtain the $\rm{L_{BLR}}$(e.g. Liu et al. 2004; Wang et al. 2006; Xiong et al. 2014). Our broad line data are collected from Xiong et al. (2014). Xiong et al. (2014) collected the largest group of Fermi blazars which contain the broad-line luminosity. These samples were derived by scaling several strong emission lines to the quasar template spectrum of Francis et al. (1991), and used Ly$\alpha$ as a reference. The luminosity of emission lines of the blazars in SDSS DR7 Quasar sample had been taken by Sbarrato et al. (2012). They had followed Celotti et al. (1997) to calculate the total luminosity of the broad lines. Particularly they set Ly$\alpha$ flux contribution to 100, and the relative weight of H$\alpha$, H$\beta$, MgII and CIV lines to 77, 22, 34 and 63 respectively. The total broad-line flux was fixed at 555.77. Their broad-line luminosity had been derived using these proportions. Shaw et al. (2012) reported on optical spectroscopy of 229 blazars in the Fermi 1LAC sample and Shen et al. (2011) had spectrally analyzed SDSS DR7 Quasar sample. Because our luminosity of emission lines are from Shaw et al. (2012) and Shen et al. (2011), and we used similar method of Sbarrato et al. (2012) to calculate broad-line luminosity. But, we found that some objects of broad-line luminosity are distinct from different samples with our results. The possible reasons are that using lines to calculate broad-line luminosity is different and variability also can cause the difference of them. In these sources, we used average broad-line luminosity instead. So we get the largest samples of Fermi blazars contain broad-line luminosity.

\subsection{The core luminosity of radio at 5 GHz}
In section 2.1, we get the largest samples of Fermi blazars contain broad-line luminosity. For the aim of study the correlation between broad-line luminosity and radio core luminosity. We use Fermi samples of introduced in section 2.1 to cross-correlated the Fermi samples from literatures which have the radio core flux at 5 GHz. So we get the largest Fermi samples not only contain the broad-line luminosity, but the radio core flux at 5 GHz.

For our samples, we can obtain the core flux at 5 GHz except for the following 15 samples:0133+476; 0403-132; 0420-014; 0458-020; OG 050; 0605-085; GB6 J0654+5042; B2 1324+22; B3 1417+385; 2155-152; PKS 2209+236; 2223-052; PKS 2227-08; TXS 2331+073; 2345-16. But we can get the core flux at 1.4 GHz. For these sources, we obtained their core flux at 1.4 GHz from Cooper et al. (2007). We use the relation expression $\rm{S_{\nu}\propto{\nu^{-\alpha}}}$, where $\alpha$ is the corresponding spectral index, to extrapolate the core flux at 5 GHz. We adopt $\alpha=0$ for the radio core flux. For these sources which we obtain their core flux, we adopted K-corrected by using equation:
\begin{equation}
\rm{S_{\nu}=S_{\nu}^{obs}(1+z)^{\alpha-1}}
\end{equation}
Where $\alpha$ is the corresponding spectral index. $\rm{\alpha_{R}=0.0}$ (Cheng et al. 2000). For the aim to study the relation between the broad-line luminosity and radio core luminosity, we need to calculate the core luminosity of radio at 5 GHz. Venters et al. (2009) gave the relation $\rm{L_{\nu}=4\pi d_{L}^{2}S_{\nu}}$to calculate the luminosity. Where $d_{L}$ is luminosity distance.
\begin{equation}
\rm{d_{\rm{L}}(z)=\frac{{c}}{H_{\rm{0}}}{(1+z)}\int\limits_0^{z}[\Omega_{\rm{\Lambda}}+\Omega_{\rm{m}}(1+z^{'})^{3}]^{-1/2}dz^{'}}
\end{equation}
So we can obtain the related luminosity of our sources.

\subsection{The intrinsic bolometric luminosity}
For the aim to study the correlation between the intrinsic bolometric luminosity and black hole mass. We need to obtain the bolometric luminosity and black hole mass. Nemmen et al. (2012) established a physical analogy between AGNs and $\gamma$-ray bursts (GRBs). A typical view in their work was that $\gamma$-ray luminosity of blazars and GRBs considered beaming effect. They computed intrinsic luminosity, L, for blazars by correcting observation which is $\gamma$-ray luminosity of isotropically equivalent,$\rm{L^{iso}}$, for the beaming factor $\rm{f_{b}}$, such that $\rm{L=f_{b}L^{iso}}$.  For blazars, $\rm{f_{b}}$ was estimated by $\rm{1-cos1/\Gamma}$ where $\rm{\Gamma}$ was the bulk Lorentz factor of the flow, since jet opening angle $\rm{\theta_{j}}$ of AGNs obey $\rm{\theta<1/\Gamma}$ (Jorstad et al. 2005; Pushkarev et al. 2009). Using VLBI and VLBA, Hovatta et al. (2009) and Pushkarev et al. (2009) calculated the variability Lorentz factors $\rm{\Gamma_{var}}$. The bulk Lorentz factors of Nemmen et al. (2012) were from the results of Hovatta et al. (2009) and Pushkarev et al. (2009).

Nemmen et al. (2012) gave another way to estimate the bolometric luminosity. The equation as follows:
\begin{equation}
\rm{L_{bol}^{iso}=L^{iso}+L_{syn}^{iso}}
\end{equation}
Where $\rm{L_{bol}^{iso}}$ is the bolometric luminosity, $\rm{L^{iso}}$ is $\gamma$-ray luminosity of isotropically equivalent, $\rm{L_{syn}^{iso}}$ is the isotropically equivalent luminosity of the synchrotron peak. We got $\rm{L_{syn}^{iso}}$ from Finke et al. (2013). There are nine sources that we can not obtain the luminosity of the synchrotron peak from Finke et al. (2013). These sources are as follows: PKS 0302-623; PKS 0308-611; 0716+714; OJ 451; S4 0913+39; 1424-418; 1622-253;  B2 1846+32; PKS 2005-489. For these sources, we use the relation was given by Meyer et al. (2011) to calculate the luminosity of the synchrotron peak. The equation as follows:
\begin{equation}
\begin{split}
\rm{{LogL_{peak}=0.61(\pm0.01)(LogL_{5 GHZ}-43)}}\\
{+45.68(\pm0.02)erg\cdot s^{-1}}
\end{split}
\end{equation}
Where $\rm{LogL_{5 GHZ}}$ is the total luminosity of 5 GHz. We computed the intrinsic bolometric luminosity as $\rm{L_{bol}=f_{b}L_{bol}^{iso}}$, using the opening angle or beaming factor to correct $\rm{L_{bol}^{iso}}$. Because of estimating the $\rm{f_{b}}$ for blazars rely on the bulk Lorentz factors $\Gamma$ which are not available for all blazars. So in the samples of Nemmen et al. (2012), they used the anti-correlation between $\rm{L_{bol}^{iso}}$ and $\rm{f_{b}}$ as an estimator of $\rm{f_{b}}$ for the blazars without measurements of $\Gamma$. So for our samples, we use the similar way to calculate the intrinsic bolometric luminosity. We obtained the $\rm{L^{iso}}$ from Nemmen et al. (2012). If we can not found the $\rm{L^{iso}}$ from Nemmen et al. (2012), we used $\rm{L_{\nu}=4\pi d_{L}^{2}S_{\nu}}$ to calculate the $\rm{L^{iso}}$.

\subsection{Black hole mass and accretion rates}
Some authors used traditional virial method to estimate the black hole mass for FSRQs(e.g. Woo \& Urry 2002; Wang et al. 2004; Liu et al. 2006; Sbarrato et al. 2012; Chai et al. 2012; Shaw et al. 2012; Shen et al. 2011). We get black hole masses from Xiong et al. (2014). For the BL Lac objects, the black hole masses can be estimated from the properties of their host galaxies with either $\rm{M_{BH}-\sigma}$ or $\rm{M_{BH}-L}$ relations. Where $\sigma$ and L are the stellar velocity dispersion and the bulge luminosity of the host galaxies (e.g. Woo \& Urry 2002; Zhou \& Cao 2009; Zhang et al. 2012; Sbarrato et al. 2012; Chai et al. 2012).

We want to study the relation between the accretion rate and the intrinsic bolometric luminosity. So we need to another parameter, accretion rates, which is an important parameter for understanding the nature of blazars. Xie et al. (2004) was based on accretion disk theory and the relativistic beaming model of blazars, used the total bolometric observed luminosity divided by $\rm{\delta^{3+\alpha}}$ and Eddington luminosity and got the intrinsic accretion rates. Ghisellini et al. (2011) and Sbarrato et al. (2012) have studied the relation between $\rm{L_{\gamma}/L_{Edd}}$ and $\rm{L_{BLR}/L_{Edd}}$. They obtained the dividing line is of $\rm{L_{BLR}/L_{Edd}\sim5\times 10^{-4}}$. Xiong et al. (2014) revisited divide line between BL Lacs and FSRQs proposed by Ghisellini et al. (2011) and Sbarrato et al. (2012). They obtained the divide line is of $\rm{L_{BLR}/L_{Edd}\sim 10^{-3}}$ corresponding to $\rm{\dot{M}/\dot{M}_{Edd}=0.1}$ ($\rm{L_{d}\approx 10L_{BLR}}$, $\rm{\dot{M}_{Edd}\equiv L_{Edd}/c^{2}}$, $\rm{ L_{d}=\eta\dot{M}c^{2}}$, $\rm{\eta=0.1}$). In addition, Ghisellini et al. (2010) studied general physical properties of bright Fermi blazars and found that there is a divide between BL Lacs and FSRQs occurring at $\rm{\dot{M}/\dot{M}_{Edd}=0.1}$. Following Ghisellini et al. (2010), Xiong et al. (2014) gave the formula about calculating the $\rm{\dot{M}/\dot{M}_{Edd}}$ as follows:
\begin{equation}
\rm{\dot{M}/\dot{M}_{Edd}\equiv \frac{\dot{M}c^{2}}{1.3\times 10^{38}(M/M_{\odot})}}
\end{equation}
For FSRQs, $\rm{\dot{M}}$ is given by $\rm{\dot{M}=L_{d}/(\eta c^{2})}$,  with $\rm{\eta=0.1}$ and for BL Lacs, $\rm{\dot{M}}$ is given by $\rm{\dot{M}=P_{jet}/c^{2}}$. We obtain the accretion rate using equation (5).

 The relevant data are listed in Table 1 with the following headings: column (1), the 2FGL name of sources; column (2), other name; column (3), classification of source: bzb=BL Lacs, baq=FSRQs; column (4), redshift of sources; column (5), the core flux density at 5 GHz in units of Jy; column (6), the references of column (5); column (7), logarithm of black hole mass of sources from Xiong et al. (2014) in units of $\rm{M_{\odot}}$; column (8), the logarithm of beaming factor $\rm{f_{b}}$  from Xiong et al. (2014); column (9), the luminosity of the synchrotron peak from Finke et al. (2013); column (10), observation $\gamma$-ray luminosity of isotropically equivalent; column (11), the logarithm of intrinsic bolometric luminosity in units of $\rm{erg\cdot s^{-1}}$; column (12), the logarithm of broad-line luminosity from Xiong et al. (2014) in units of $\rm{erg\cdot s^{-1}}$ ; column (13), the accretion rate in Eddington unit which is given by equation (5).

\section{RESULTS}
\subsection{ The distributions}
In Fig.1, we give the redshift distributions of the various classes. The redshift distributions for our samples are 0$<$z$<$3.1 and mean value is $\rm{0.80\pm0.06}$.  Mean values for FSRQs and BL Lacs are $\rm{1.05\pm0.07}$ and $\rm{0.33\pm0.05}$ respectively. We can obtain that our samples contain a wide range of redshift. For this we could exclude the selection effect of redshift.

In Fig.2, we give the distributions of radio core luminosity at 5 GHz. The radio core luminosity for our samples is$\rm{10^{40}\sim10^{45.5}erg\cdot s^{-1}}$ and mean value is $\rm{10^{43.38\pm0.13}erg\cdot s^{-1}}$. Mean values for FSRQs and BL Lacs are $\rm{10^{43.94\pm0.09}erg\cdot s^{-1}}$ and $\rm{10^{42.22\pm0.23}erg\cdot s^{-1}}$.

Fig.3 shows the distribution of broad-line luminosity for our samples. The range of our samples is $\rm{10^{41.5}\sim10^{46.5}erg\cdot s^{-1}}$. The mean value is $\rm{10^{44.51\pm0.11}erg\cdot s^{-1}}$. Mean values for FSRQs and BL Lacs are $\rm{10^{44.90\pm0.08}erg\cdot s^{-1}}$ and $\rm{10^{43.20\pm0.20}erg\cdot s^{-1}}$.  For our samples, we can see that the mean value broad-line luminosity, FSRQs is larger than that of BL Lacs.

Fig.4 shows the distribution of black hole mass for our sources. The black hole mass distributions for our samples mainly are $\rm{10^{7.4}M_{\odot}\sim10^{9.8}M_{\odot}}$ and mean value is $\rm{10^{8.63\pm0.05}M_{\odot}}$. Mean values for FSRQs and BL Lacs are $\rm{10^{8.78\pm0.05}M_{\odot}}$ and $\rm{10^{8.38\pm0.07}M_{\odot}}$ respectively. We find that the black hole mass is also contain a wide range.

Fig.5 shows the distribution of intrinsic bolometric luminosity for our samples, which the distributions mainly are $\rm{10^{42}\sim10^{48.8}erg\cdot s^{-1}}$ and the mean value is $\rm{10^{45.36\pm0.13}erg\cdot s^{-1}}$. Mean values for FSRQs and BL Lacs are $\rm{10^{45.70\pm0.16}erg\cdot s^{-1}}$ and $\rm{10^{44.72\pm0.19}erg\cdot s^{-1}}$. We can obtain that the mean intrinsic bolometric luminosity, FSRQs is larger than that of BL Lacs.

\subsection{Broad-line luminosity vs radio core luminosity at 5 GHz}
The broad-line luminosity can be taken as an indicator of the accretion power of the source (Celotti et al. 1997). The radio core luminosity was believed to be a straightforward indicator of jet power (Blandford \& Konigl 1979). We study the correlation between broad-line emission and radio core emission for our samples which detected by Fermi LAT.

Using the data of Table 1 and adopting the ordinary least-squares(OLS) method to fitting the relation between $\rm{L_{BLR}}$ and $\rm{L_{R}^{C}}$, we give the result of linear regression analysis as follows:
\begin{equation}
\rm{LogL_{BLR}=(0.81\pm0.06)L_{R}^{C}+(9.17\pm2.93)}
\end{equation}
With a correlation coefficient r=0.79 and a chance possibility p$<$0.0001; the slope of the linear regression equation is $\rm{0.81\pm0.06}$ and close to 1. Fig.6 shows the relation between broad-line luminosity and the radio core luminosity. As a comparison to the result of linear regression, we also using the OLS method to analysis the correlation between broad-line emission and radio core emission for BL Lacs and FSRQs separately. With correlation coefficient r=0.89 for BL Lacs and r=0.69 for FSRQs, chance possibility p$<$0.0001; the slope of the linear regression equation are $\rm{0.64\pm0.07}$ for BL Lacs and $\rm{0.60\pm0.08}$ for FSRQs. It is well known that the use of luminosity instead of flux often introduces a redshift bias to the data. As point out by Padovani (1992), the luminosity was strongly correlated with redshift and would lead a spurious correlation. So we adopt a Pearson partial correlation analysis in our analysis to exclude the redshift effect. The linear correlation coefficient between $\rm{LogL_{BLR}}$ and ${\rm L_{R}^{C}}$ with the effect of redshift exclude is r=0.58 and a chance probability of $\rm{p=3.17\times 10^{-9}}$. The linear correlation coefficient between $\rm{LogL_{BLR}}$ and ${\rm L_{R}^{C}}$ with the effect of redshift exclude are r=0.69 and a chance probability of $\rm{p=7.70\times 10^{-4}}$ for BL Lacs and r=0.49 and a chance probability of $\rm{p=1.88\times 10^{-5}}$ for FSRQs. The correlation between broad-line and radio core emission suggests a close link between the formation of jets and accretion onto the central Black Hole in our work.

\subsection{The Black Hole Mass vs the intrinsic bolometric luminosity}
The intrinsic bolometric luminosity also is an important parameter for understanding the nature of blazars. Xie et al. (2004) was based on the variation timescale estimated the black hole mass and used $\rm{\delta^{3+\alpha}}$ to exclude the beaming effect. Where $\rm{\delta}$ is the Doppler factor and $\rm{\alpha}$ is the spectral index. Because of observing bolometric luminosity must be corrected for the Doppler effect. So Xie et al. (2004) used the observed luminosity to divide the $\rm{\delta^{3+\alpha}}$ to exclude the beaming effect and studied the mass-luminosity relation.

In this part, we use another theory to obtain the intrinsic bolometric luminosity in section 2.3. in section 2.4, we give the approach to obtain the black hole mass and study the relation between the intrinsic bolometric luminosity and the black hole mass. Using the OLS method to fitting the relation between the intrinsic bolometric luminosity and the black hole mass. Fig.7 shows the relation between the intrinsic bolometric luminosity and the black hole mass. The result of linear regression analysis as follows:
\begin{equation}
\rm{{Log}\frac{L_{in}}{L_{\odot}}=(0.95\pm0.26){Log}\frac{M}{M_{\odot}}+(3.53\pm2.24)}
\end{equation}
The Pearson correlation is r=0.337 for and probability is $\rm{p=3.882\times 10^{-4}}$. The analysis results indicate that there is high significant correlation between intrinsic bolometric luminosity and the black hole mass. From Fig.7, we can see that the data points have large scatter. For this, we also use the Spearman correlate analysis. The Spearman correlation is r=0.382 for and probability is $\rm{p=4.892\times 10^{-5}}$. The analysis results indicate that the correlation between intrinsic bolometric luminosity and the black hole mass really exist. This means that there is a significant correlation between the black hole mass and the intrinsic bolometric luminosity in FSRQs and BL Lacs of our Fermi samples. In addition, we can obtain that FSRQs occupy the region of high luminosity and larger mass and BL Lacs occupy the region of low luminosity and smaller mass while some FSRQs and BL Lacs occupy the region of intermediate luminosity and mass.

From equation (7), we have
\begin{equation}
\rm{\frac{L_{in}}{L_{\odot}}\propto(\frac{M}{M_{\odot}})^{0.95\pm0.26}}
\end{equation}
We can observe that the mass-luminosity relation for Fermi blazars derived in this work is a powerlaw relation similar to that for main-sequence stars.

\subsection{The accretion rates vs the intrinsic bolometric luminosity}
 The accretion rate is an important parameter for understanding the nature of blazars. Xie et al. (2004) based on accretion disk theory and the relativistic beaming model of blazars, used the total bolometric observed luminosity divided by $\rm{\delta^{3+\alpha}}$ and Eddington luminosity estimated the accretion rates. In section 2.4, we give the way how to estimate the accretion rates in this work. So we recalculate the accretion rates for our samples. From Table 1 of column (13), we give the accretion rates. From Table 1 and Fig. 8, we can obtain that the accretion rates of FSRQs and BL Lacs are quite different for our Fermi blazars. FSRQs have high luminosity and higher accretion rates than that of BL Lacs. The same trend has also been found by Xie et al. (2004) and O'Dowd et al. (2002) in their research.

\section{DISCUSSIONS}
\subsection{The Broad-line luminosity and radio core luminosity}
From our results of in section 3.2, we can see that the correlation between broad-line luminosity and radio core luminosity is significant which supports that jets has a close link with accretion onto the central BH. According to Ghisellini (2006), if relativistic jets are powered by a Poynting flux, under some reasonable assumptions, the Blandford \& Znajek (1977) power can be written as:
\begin{equation}
\rm{L_{BZ}\sim 6\times 10^{20}(\frac{a}{m})^{2}(\frac{M_{BH}}{M_{\odot}})^{2}B^{2}erg\cdot s^{-1}}
\end{equation}
Where $\rm{L_{BZ}}$  is the luminosity, $\rm{\frac{a}{m}}$ is the specific black hole angular momentum ($\sim$ 1 for a maximally rotating BH), and B is the magnetic field in Gauss. Assuming that the value of the magnetic energy density $\rm{U_{B}\equiv \frac{B^{2}}{8\pi}}$, close to the black hole which is a fraction $\rm{\varepsilon_{B}}$ of the available gravitational energy:
\begin{equation}
\rm{U_{B}=\varepsilon_{B}\frac{GM_{BH}\rho}{R}=\varepsilon_{B}\frac{R_{S}\rho c^{2}}{2R}}
\end{equation}
Where $\rm{R_{S}=2GM_{BH}/c^{2}}$ is Schwarzschild radius, $\rho$ is the mass density, R is the radius, c is the speed of light. The density $\rho$ is linked to the accretion rate $\rm{\dot{M}}$ through
\begin{equation}
\rm{\dot{M}=2\pi RH\rho \beta_{r}c}
\end{equation}
Where $\rm{\beta_{r}c}$ is the radial infalling velocity, H is the disk thickness. The mass accretion rate $\rm{\dot{M}}$ is linked to the observed luminosity produced by the disk
\begin{equation}
\rm{L_{disk}=\eta \dot{M}c^{2}}
\end{equation}
So from equations (9)-(12), The BZ jet power can then be written as:
\begin{equation}
\rm{L_{BZ,jet}\sim (\frac{a}{m})^{2}\frac{R_{S}^{3}}{HR^{2}}\frac{\varepsilon_{B}}{\eta}\frac{L_{disk}}{\beta_{r}}}
\end{equation}
Ghisellini (2006) gave the maximum BZ jet power and can be written as:
\begin{equation}
\rm{L_{jet}\sim \frac{L_{disk}}{\eta}}
\end{equation}
In addition, based on the current theories of accretion disks, the BLR is ionized by the accretion disk. So we have
\begin{equation}
\rm{L_{disk}\approx 10L_{BLR}}
\end{equation}
Based on the equations (14) and (15), we have
\begin{equation}
\rm{L_{BLR}\sim 0.1\eta L_{jet}}
\end{equation}
From equation (16), we can have
\begin{equation}
\rm{LogL_{BLR}=LogL_{jet}+Log\eta+const}
\end{equation}
From equation (17), we find that the theoretical predicted coefficient of the $\rm{LogL_{BLR}\sim LogL_{jet}}$ relation is 1. Using linear regression analysis, we have $\rm{LogL_{BLR}\sim (0.81\pm0.06)L_{R}^{C}}$ for our Fermi blazars, which is consistent with the theoretical predicted coefficient of the $\rm{LogL_{BLR}\sim LogL_{jet}}$ relation. So our results suggest that Fermi blazars jets are also powered by energy extraction from a rapidly spinning black hole through the magnetic field provided by the accretion disk. This supports the hypothesis provided by Xie et al. (2007). Compared the results with Xie et al. (2007), and we have a lager samples and all blazar sources detected by Fermi LAT. Moreover we notice that $\rm{LogL_{BLR}\sim (0.64\pm0.07)L_{R}^{C}}$ for BL Lacs and $\rm{LogL_{BLR}\sim (0.60\pm0.08)L_{R}^{C}}$ for FSRQs. The slope of the linear regression equations are $\rm{0.64\pm0.07}$ for BL Lacs and $\rm{0.60\pm0.08}$ for FSRQs and not close to 1. The possible reasons are that there is a clear segregation in BLR luminosity betweenthe two classes or the samples are not complete. At last, we should be caution in the core luminosity at 5 GHz. The rapid variability and the beaming effect may bring in uncertainties. For broad line luminosity, it also is worth noting that BLR luminosity is not a direct measure of the disc luminosity and furthermore that it is possible that BL Lacs may have a less luminous accretion disc (radiatively inefficient accretion flow) than FSRQs which would make it difficult to estimate the accretion rate from the BLR or disc luminosity.

\subsection{The mass-luminosity relation and the accretion rate}
 In section 3.3, we give the relation between black hole mass and the intrinsic bolometric luminosity. Based on the linear regression analysis, we obtain the relation of equation (7). We also give the mass-luminosity relation of equation (8). We obtain that there is a significant correlation between the black hole mass and the intrinsic bolometric luminosity in FSRQs and BL Lacs of our Fermi samples. Our results are consistent with Xie et al. (2004). Xie et al. (2004) was based on the variation timescale estimated the black hole mass and used $\rm{\delta^{3+\alpha}}$ to exclude the beaming effect. But the variation timescale and the $\delta$ can not obtain for most samples. So in this work, we adopt another way to obtain the intrinsic bolometric luminosity and using another factor $\rm{f_{b}}$ to exclude the beaming effect. Compared with Xie et al. (2004), for our results, from Fig. 7, we can see that the data points have large scatter. But also have significant correlation. For the data points have large scatter, the probably reason is that we obtain the black hole mass, intrinsic bolometric luminosity and using the beaming factor are different from Xie et al. (2004). But we have a lager samples and all sources detected by Fermi LAT. From equation (8), we can observe that the mass-luminosity relation for Fermi blazars derived in this work is a powerlaw relation similar to that for main-sequence stars. We should be caution in the black hole mass estimated from virial method or relations of host galaxies because of non-thermal dominance for blazars and contamination of host galaxy light.

 Xie et al. (2004) was based on accretion disk theory and the relativistic beaming model of blazars, used the total bolometric observed luminosity divided by $\rm{\delta^{3+\alpha}}$ and Eddington luminosity and got the intrinsic accretion rates. In section 3.4, we study the relation between accretion rates and the intrinsic bolometric luminosity. We use our samples and recalculate the accretion rates using the method of mentioned in section 2.4. We obtain the similar results with Xie et al. (2004). For FSRQs the accretion rates are larger. A possible interpretation of this phenomenon is the exhaustion of the gas stockpile available for accretion in the host galaxy (Cattaneo et al. 1999; Haehnelt et al. 2000). Cattaneo et al. (1999) and Haehnelt et al. (2000) argued that with the gradual depletion of the gas, FSRQs (mainly fueled by accretion) will switch to BL Lac objects (mostly fed by Kerr black hole rotational energy extraction via the BZ mechanism). However, for intrinsic bolometric luminosity, the variability and beaming factor estimated can lead to uncertainty.

 In this paper, our results match that the theoretical/observational accretion rates proposed in the dual jet theories (Punsly 1996a; 1996b). In the dual jet theories, the evolutionary sequence from FSRQs to BL Lac objects. (Punsly 1996a; 1996b). Vagnetti et al. (1991). pointed out that accretion rate is the most strongly evolving parameter in a unified scheme for AGNs. The relation between the accretion rate and the intrinsic bolometric luminosity also show that FSRQs occur in the earlier, violent phase of the blazar evolutionary sequence, while BL Lac objects occur in the late. This evidence suggests that the evolutionary track of AGNs is from FSRQs to BL Lac objects. This result is consistent with Xie et al. (2004).  So the accretion rate is an important parameter for understanding the nature of blazars.

\section{CONCLUSIONS}
In this work we have analyzed a large sample of blazars detected by Fermi LAT. Our main results are the following:

(1) We study the relation between broad-line luminosity and radio core luminosity at 5 GHz. A strong correlation between broad-line luminosity and radio core luminosity is found for our samples detected by Fermi LAT suggest that a close link between the formation of jets and accretion onto the central Black Hole.we have $\rm{LogL_{BLR}\sim (0.81\pm0.06)L_{R}^{C}}$ for our samples. The slope of our results is $0.81\pm0.06$ , which is very close to 1. There is a significant correlation between broad-line luminosity and radio core luminosity, with the effect of redshift exclude, r=0.58 and a chance probability of $\rm{p=3.17\times 10^{-9}}$. Our results provide support for the theories of jet-disk symbiosis of Ghisellini (2006) and Maraschi \& Tavecchio (2003). From Fig.6, we can see that BL Lac objects appear to lie on the same linear correlation with powerful quasars, and supporting the view that these two classes of blazars have similar jets. Our results support the unification of the two classes of sources into a single Fermi blazar population.

(2) The correlation between the intrinsic bolometric luminosity and the black hole mass which supports mass-luminosity relation for Fermi blazars derived in this work is a powerlaw relation similar to that for main-sequence stars. The result of linear regression analysis between the intrinsic bolometric luminosity and the black hole mass give the Pearson correlation is r=0.337 for and probability is $\rm{p=3.882\times10^{-4}}$. We also use the Spearman correlate analysis. The Spearman correlation is r=0.382 for and probability is $\rm{p=4.892\times10^{-5}}$. This means that there is a significant correlation between the black hole mass and the intrinsic bolometric luminosity in FSRQs and BL Lacs of our Fermi samples. In addition, we can obtain that FSRQs occupy the region of high luminosity and larger mass and BL Lacs occupy the region of low luminosity and smaller mass while some FSRQs and BL Lacs occupy the region of intermediate luminosity and mass.

(3) We study the relation between accretion rates and the intrinsic bolometric luminosity. We obtain that the accretion rates of FSRQs and BL Lacs are quite different for our Fermi blazars. FSRQs have high luminosity and higher accretion rates than that of BL Lacs. For FSRQs the accretion rates are larger. A possible interpretation of this phenomenon is the exhaustion of the gas stockpile available for accretion in the host galaxy (Cattaneo et al. 1999; Haehnelt et al. 2000). We obtain that the evolutionary track of AGNs is from FSRQs to BL Lac objects.

\acknowledgments We thank the anonymous referee for valuable comments and suggestions. We are very grateful to the Science Foundation of Yunnan Province of China(2012FB140,2010CD046). This work is supported by the National Nature Science Foundation of China (11063004, 11163007,U1231203), and the High-Energy Astrophysics Science and Technology Innovation Team of Yunnan Higher School and Yunnan Gravitation Theory Innovation Team (2011c1). This research has made use of the NASA/IPAC Extragalactic Database (NED), that is operated by Jet Propulsion Laboratory, California Institute of Technology, under contract with the National Aeronautics and Space Administration.

\clearpage
\begin{figure}
\epsscale{}
 \plotone{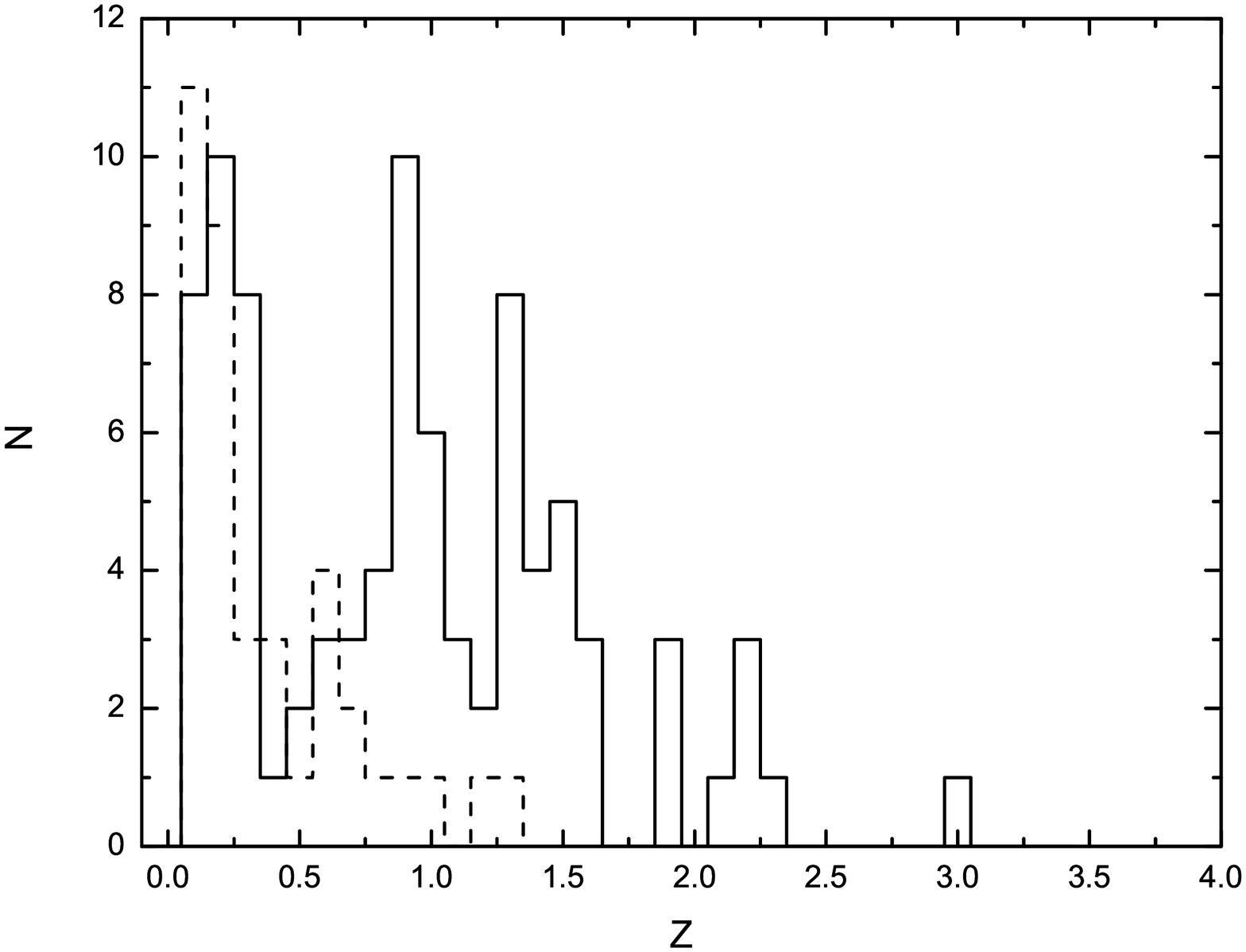}
 \caption{Distributions of redshift for Fermi blazars.BL Lacs(dash line), FSRQs(solid line)}
\end{figure}

\begin{figure}
\epsscale{}
 \plotone{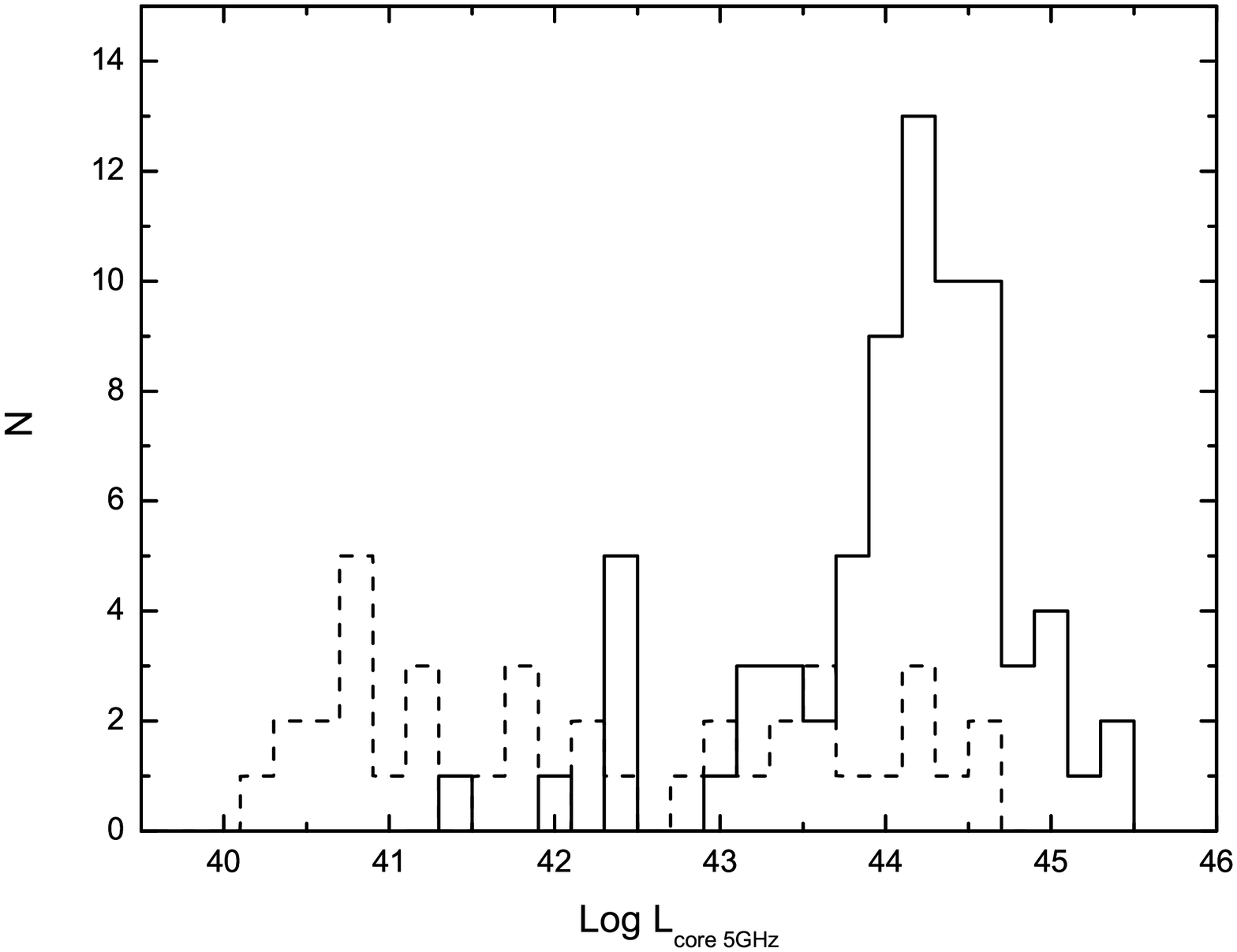}
 \caption{Distributions of core luminosity at 5 GHz($\rm{L_{R}^{C}}$) for Fermi blazars. The meanings of different lines are as same as Fig.1.}
\end{figure}

\begin{figure}
\epsscale{}
 \plotone{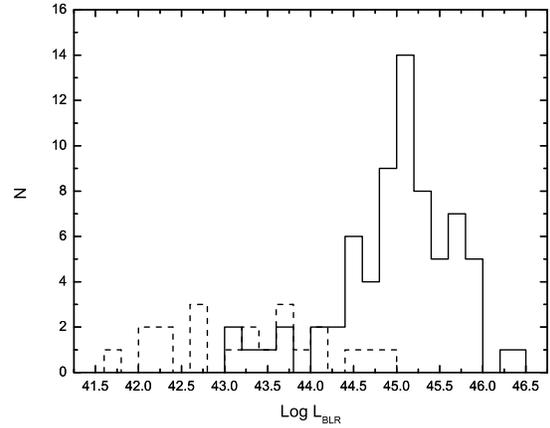}
 \caption{Distributions of broad-line luminosity ($\rm{L_{BLR}}$) for Fermi blazars.The meanings of different lines are as same as Fig.1.}
\end{figure}

\begin{figure}
\epsscale{}
 \plotone{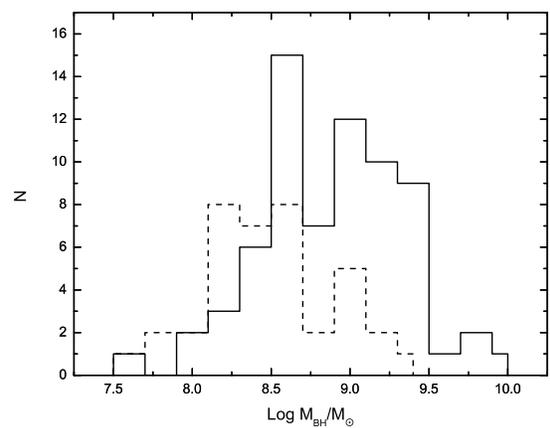}
 \caption{Distributions of Black Hole Mass for Fermi blazars in units of $\rm{M_{\odot}}$. The meanings of different lines are as same as Fig.1.}
\end{figure}

\begin{figure}
\epsscale{}
 \plotone{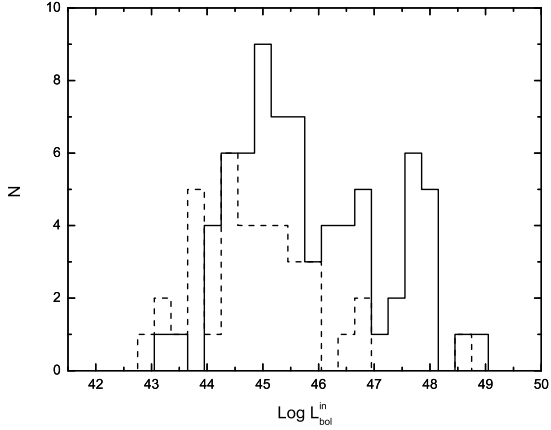}
 \caption{Distributions of intrinsic bolometric luminosity for Fermi blazars. The meanings of different lines are as same as Fig.1.}
\end{figure}

\begin{figure}
\epsscale{}
 \plotone{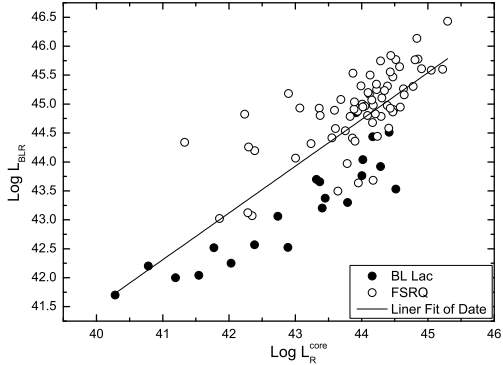}
 \caption{The relation between broad-line luminosity and the radio core luminosity for Fermi blazars. We give the result of linear regression $\rm{LogL_{BLR}=(0.81\pm0.06)L_{R}^{C}+(9.17\pm2.93)}$ and the slope of the linear regression equation is $\rm{0.81\pm0.06}$ and close to 1. BL Lacs(filled points), FSRQs(open circles). The black line is the result of linear regression.}
\end{figure}

\begin{figure}
\epsscale{}
 \plotone{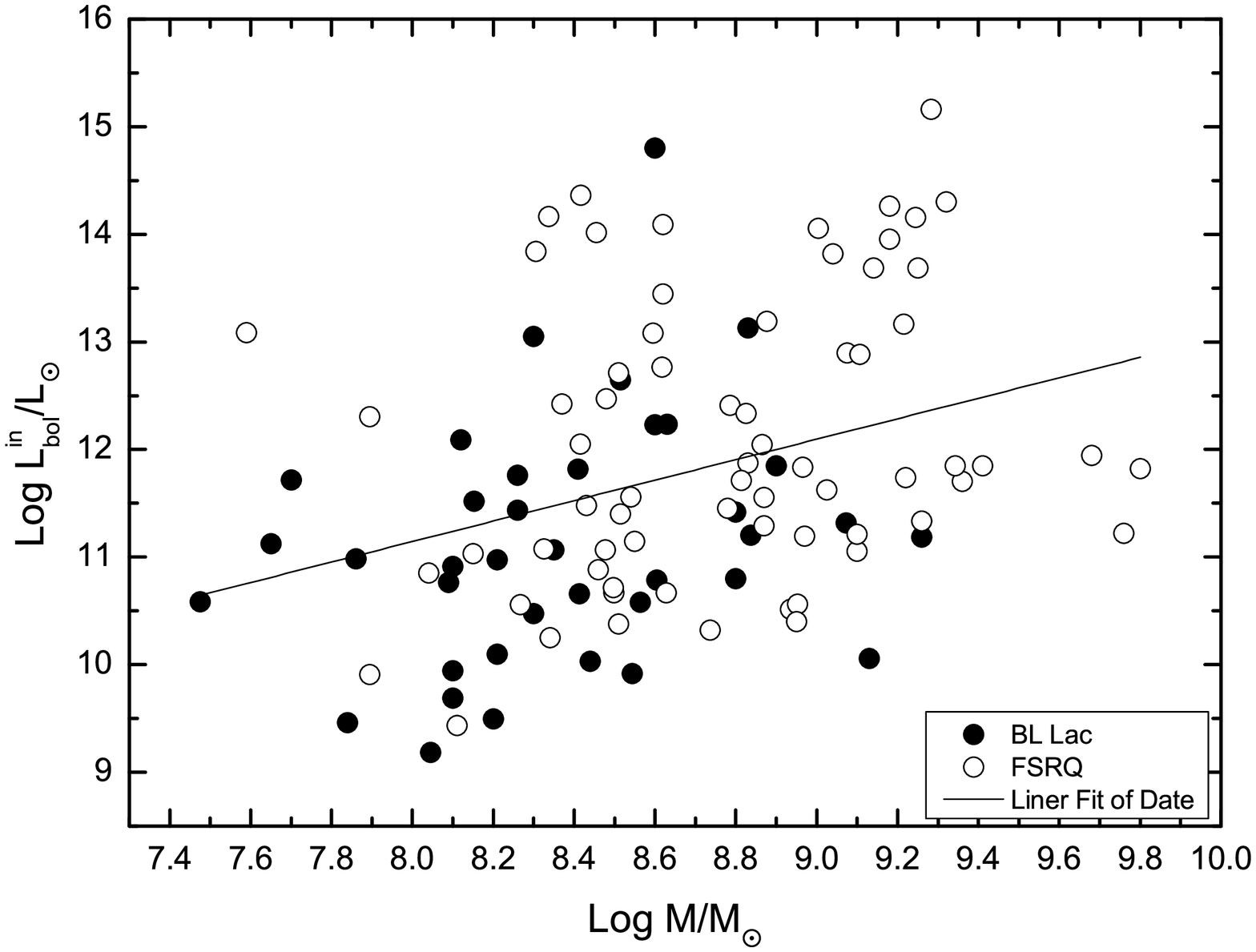}
 \caption{The relation between the intrinsic bolometric luminosity and the black hole mass. We give the result of linear regression$\rm{{Log}\frac{L_{in}}{L_{\odot}}=(0.95\pm0.26){Log}\frac{M}{M_{\odot}}+(3.53\pm2.24)}$. The Pearson correlation is r=0.337. The Spearman correlation is r=0.382. The meanings of different symbols and line are as same as Fig.6. }
\end{figure}

\begin{figure}
\epsscale{}
 \plotone{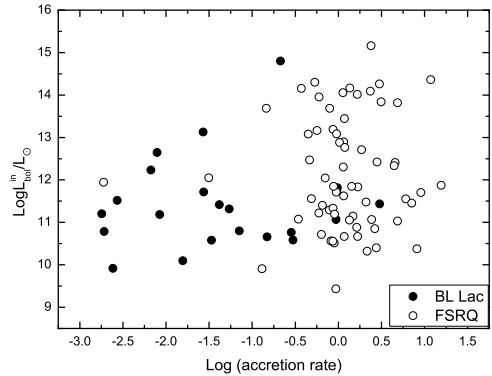}
 \caption{The relation between the intrinsic bolometric luminosity and the accretion rates. The accretion rates of FSRQs and BL Lacs are quite different for our Fermi blazars. FSRQs have high luminosity and higher accretion rates than that of BL Lacs.The meanings of different symbols are as same as Fig.6. }
\end{figure}

\clearpage
\begin{deluxetable}{crccccccccccccccccrl}
\tabletypesize{\scriptsize} \rotate \tablecaption{The sample of 111 Fermi blazars\label{tbl-1}} \tablewidth{0pt}
\tablehead{\colhead{Fermi Name} & \colhead{other name} & \colhead{Type} & \colhead{Redshift}& \colhead{$Flux_{\rm{5 GHzcore}}$} & \colhead{Ref} & \colhead{$\rm{Log{M_{BH}}}$} & \colhead{$\rm{Logf_{b}}$} & \colhead{$\rm{LogL_{peak}}$}  & \colhead{$\rm{LogL^{iso}}$}
& \colhead{$\rm{LogL_{bol}^{in}}$} & \colhead{$\rm{LogL_{BLR}}$} & \colhead{$\rm{Log\frac{\dot{M}}{\dot{M}_{Edd}}}$}\\
\colhead{(1)} & \colhead{(2)} & \colhead{(3)} & \colhead{(4)} &
\colhead{(5)} & \colhead{(6)} & \colhead{(7)} & \colhead{(8)} &
\colhead{(9)} & \colhead{(10)} & \colhead{(11)} & \colhead{(12)} & \colhead{(13)} } \startdata
2FGLJ0108.6+0135	&	0106+013	&	bzq	&	2.107 	&	1.35	&	1	&	8.83 	&	-3.21 	&	47.50 	&	48.64 	&	45.46 	&	46.13 	&	1.191 	\\
2FGLJ0113.7+4948	&	0110+495	&	bzq	&	0.389 	&	0.68	&	2	&	8.34 	&	-2.03 	&	45.25 	&	45.75 	&	43.84 	&		&		\\
2FGLJ0136.9+4751	&	0133+476	&	bzq	&	0.859 	&	1.882	&	3	&	8.52 	&	-2.62 	&	46.50 	&	47.57 	&	44.99 	&	44.44 	&	-0.185 	\\
2FGLJ0237.8+2846	&	0234+285	&	bzq	&	1.213 	&	0.54	&	1	&	9.22 	&	-2.52 	&	46.98 	&	47.78 	&	45.32 	&	45.31 	&	-0.019 	\\
2FGLJ0238.7+1637	&	0235+164	&	bzb	&	0.940 	&	1.75	&	4	&	9.07 	&	-2.66 	&	47.09 	&	47.38 	&	44.90 	&	43.92 	&	-1.267 	\\
2FGLJ0259.5+0740	&	0256+075	&	bzq	&	0.893 	&	0.44	&	1	&		&	-2.40 	&	46.10 	&	46.62 	&	46.73 	&	43.50 	&		\\
2FGLJ0303.5-6209	&	PKS 0302-623	&	bzq	&	1.348 	&	1.68	&	1	&	9.76 	&	-2.68 	&	46.72 	&	47.40 	&	44.81 	&	45.65 	&	-0.227 	\\
2FGLJ0310.0-6058	&	PKS 0308-611	&	bzq	&	1.479 	&	0.33	&	1	&	8.87 	&	-2.75 	&	46.62 	&	47.58 	&	44.88 	&	44.88 	&	-0.105 	\\
2FGLJ0339.4-0144	&	0336-019	&	bzq	&	0.852 	&	1.13	&	1	&	8.94 	&	-3.03 	&	46.64 	&	46.96 	&	44.10 	&	45.00 	&	-0.049 	\\
2FGLJ0405.8-1309	&	0403-132	&	bzq	&	0.571 	&	4.188	&	3	&	9.08 	&	-2.17 	&	46.33 	&	45.95 	&	46.48 	&	45.25 	&	0.061 	\\
2FGLJ0416.7-1849	&	0414-189	&	bzq	&	1.536 	&	0.2	&	1	&		&	-2.66 	&	46.96 	&	47.37 	&	44.84 	&	44.54 	&		\\
2FGLJ0423.2-0120	&	0420-014	&	bzq	&	0.916 	&	2.898	&	3	&	8.81 	&	-2.40 	&	46.91 	&	47.62 	&	45.30 	&	44.90 	&	-0.026 	\\
2FGLJ0428.6-3756	&	PKS0426-380	&	bzb	&	1.111 	&	0.68	&	1	&	8.60 	&	-3.00 	&	46.92 	&	48.37 	&	48.38 	&	44.04 	&	-0.673 	\\
2FGLJ0442.7-0017	&	0440?003	&	bzq	&	0.844 	&	2.21	&	1	&	8.46 	&	-2.72 	&	46.47 	&	47.57 	&	47.60 	&	44.79 	&	0.219 	\\
2FGLJ0501.2-0155	&	0458-020	&	bzq	&	2.291 	&	1.029	&	3	&	8.97 	&	-2.74 	&	47.59 	&	48.02 	&	45.42 	&	45.30 	&	0.225 	\\
2FGLJ0530.8+1333	&	0528+134	&	bzq	&	2.060 	&	0.23	&	1	&	9.80 	&	-2.96 	&	47.36 	&	48.32 	&	45.41 	&		&		\\
2FGLJ0532.7+0733	&	OG 050	&	bzq	&	1.254 	&	1.532	&	3	&	8.43 	&	-2.80 	&	47.12 	&	47.78 	&	45.07 	&	44.86 	&	0.320 	\\
2FGLJ0608.0-0836	&	0605-085	&	bzq	&	0.872 	&	1.184	&	3	&	8.63 	&	-3.01 	&	46.88 	&	47.03 	&	44.25 	&	44.97 	&	0.224 	\\
2FGLJ0654.5+5043	&	GB6 J0654+5042	&	bzq	&	1.253 	&	0.3133	&	5	&	8.33 	&	-1.90 	&	46.54 	&	45.17 	&	44.66 	&	43.97 	&	-0.465 	\\
2FGLJ0710.5+5908	&	0706+591	&	bzb	&	0.125 	&	0.0338	&	6	&	8.26 	&	-1.74 	&	45.24 	&	44.70 	&	45.35 	&		&		\\
2FGLJ0721.9+7120	&	0716+714	&	bzb	&	0.300 	&	0.6	&	2	&	8.10 	&	-2.32 	&	45.77 	&	46.78 	&	44.50 	&		&		\\
2FGLJ0738.0+1742	&	0735+178	&	bzb	&	0.424 	&	1.29	&	2	&	8.30 	&	-2.38 	&	45.80 	&	46.57 	&	46.64 	&		&		\\
2FGLJ0739.2+0138	&	0736+017	&	bzq	&	0.189 	&	0.6	&	1	&	8.11 	&	-2.74 	&	45.10 	&	45.65 	&	43.02 	&	44.19 	&	-0.029 	\\
2FGLJ0746.6+2549	&	B2 0743+25	&	bzq	&	2.979 	&	0.325	&	2	&	9.41 	&	-3.11 	&	47.37 	&	48.52 	&	45.43 	&	45.47 	&	-0.056 	\\
2FGLJ0747.7+4501	&	B3 0745+453	&	bzq	&	0.192 	&	0.051	&	2	&	8.54 	&	-1.77 	&	44.90 	&	44.78 	&	45.14 	&	44.34 	&	-0.314 	\\
2FGLJ0750.6+1230	&	0748+126	&	bzq	&	0.889 	&	1.06	&	1	&	8.15 	&	-2.52 	&	46.58 	&	46.99 	&	44.62 	&	44.95 	&	0.686 	\\
2FGLJ0757.1+0957	&	0754+100	&	bzb	&	0.266 	&	1.25	&	2	&	8.20 	&	-2.97 	&	45.78 	&	45.72 	&	43.08 	&		&		\\
2FGLJ0809.8+5218	&	1ES 0806+524	&	bzb	&	0.138 	&	0.066	&	6	&	8.90 	&	-1.89 	&	45.13 	&	45.14 	&	45.43 	&		&		\\
2FGLJ0824.7+3914	&	4C +39.23	&	bzq	&	1.216 	&	0.88	&	2	&	8.55 	&	-2.51 	&	46.90 	&	46.98 	&	44.73 	&	44.83 	&	0.170 	\\
2FGLJ0824.9+5552	&	OJ 535	&	bzq	&	1.418 	&	1	&	2	&	9.26 	&	-2.73 	&	47.03 	&	47.53 	&	44.92 	&	45.31 	&	-0.063 	\\
2FGLJ0825.9+0308	&	0823+033	&	bzb	&	0.506 	&	0.9	&	1	&	8.83 	&	-2.14 	&	46.65 	&	45.86 	&	46.72 	&	43.37 	&	-1.570 	\\
2FGLJ0830.5+2407	&	B2 0827+24	&	bzq	&	0.940 	&	1.359	&	7	&	8.95 	&	-3.10 	&	46.45 	&	47.17 	&	44.15 	&	44.98 	&	-0.086 	\\
2FGLJ0831.9+0429	&	PKS 0829+046	&	bzb	&	0.174 	&	0.71	&	1	&	8.63 	&	-2.08 	&	45.25 	&	45.68 	&	45.82 	&	42.57 	&	-2.176 	\\
2FGLJ0834.3+4221	&	OJ 451	&	bzq	&	0.249 	&	0.31	&	2	&	9.68 	&	-1.93 	&	45.22 	&	45.24 	&	45.53 	&	43.07 	&	-2.724 	\\
2FGLJ0841.6+7052	&	0836+710	&	bzq	&	2.172 	&	3.743	&	2	&	9.36 	&	-3.20 	&	47.90 	&	48.36 	&	45.29 	&	46.43 	&	0.956 	\\
2FGLJ0854.8+2005	&	0851+202	&	bzb	&	0.306 	&	2.3	&	2	&	8.56 	&	-2.23 	&	46.06 	&	46.12 	&	44.16 	&	43.21 	&	-1.472 	\\
2FGLJ0903.4+4651	&	S4 0859+470	&	bzq	&	1.466 	&	1.645	&	2	&	9.25 	&	-2.51 	&	46.99 	&	46.95 	&	47.27 	&	45.26 	&	-0.100 	\\
2FGLJ0909.1+0121	&	0906+015	&	bzq	&	1.024 	&	1.002	&	2	&	9.00 	&	-2.74 	&	46.59 	&	47.60 	&	47.64 	&	45.17 	&	0.052 	\\
2FGLJ0917.0+3900	&	S4 0913+39	&	bzq	&	1.267 	&	0.62	&	2	&	8.62 	&	-1.93 	&	46.27 	&	46.95 	&	47.03 	&	44.80 	&	0.070 	\\
2FGLJ0920.9+4441	&	B3 0917+449	&	bzq	&	2.190 	&	1.311	&	2	&	9.28 	&	-3.12 	&	47.49 	&	48.72 	&	48.74 	&	45.78 	&	0.380 	\\
2FGLJ0937.6+5009	&	CGRaBS J0937+5008	&	bzq	&	0.275 	&	0.217	&	2	&	7.90 	&	-1.84 	&	44.57 	&	45.25 	&	43.49 	&	43.12 	&	-0.885 	\\
2FGLJ0948.8+4040	&	4C +40.24	&	bzq	&	1.249 	&	0.7	&	1	&	8.95 	&	-3.28 	&	46.95 	&	46.98 	&	43.98 	&	45.50 	&	0.438 	\\
2FGLJ0956.9+2516	&	0953+254	&	bzq	&	0.707 	&	0.37	&	1	&	8.88 	&	-2.41 	&	46.11 	&	46.67 	&	46.78 	&	44.93 	&	-0.062 	\\
2FGLJ0957.7+5522	&	4C+55.17	&	bzq	&	0.896 	&	2.568	&	2	&	8.34 	&	-2.77 	&	46.78 	&	47.70 	&	47.75 	&	44.58 	&	0.131 	\\
2FGLJ0958.6+6533	&	0954+658	&	bzb	&	0.368 	&	0.477	&	4	&	8.52 	&	-2.18 	&	45.88 	&	45.98 	&	46.23 	&	42.53 	&	-2.104 	\\
2FGLJ1001.0+2913	&	GB6 J1001+2911	&	bzb	&	0.558 	&	0.142	&	2	&	7.48 	&	-2.21 	&	45.90 	&	46.20 	&	44.17 	&	43.06 	&	-0.526 	\\
2FGLJ1014.1+2306	&	4C+23.24	&	bzq	&	0.566 	&	0.99	&	1	&	8.51 	&	-2.21 	&	45.91 	&	46.07 	&	46.30 	&	44.89 	&	0.270 	\\
2FGLJ1015.1+4925	&	1ES 1011+496	&	bzb	&	0.212 	&	0.242	&	2	&	8.30 	&	-2.11 	&	45.78 	&	45.93 	&	44.06 	&		&		\\
2FGLJ1017.0+3531	&	B2 1015+35	&	bzq	&	1.228 	&	0.9	&	2	&	9.10 	&	-2.54 	&	46.62 	&	47.04 	&	44.64 	&	45.34 	&	0.129 	\\
2FGLJ1043.1+2404	&	B2 1040+024	&	bzb	&	0.559 	&	0.609	&	2	&	8.09 	&	-2.18 	&	46.31 	&	46.13 	&	44.35 	&	43.66 	&	-0.547 	\\
2FGLJ1057.0-8004	&	PKS1057-79	&	bzb	&	0.581 	&	2.43	&	1	&	8.80 	&	-2.33 	&	46.28 	&	46.52 	&	44.38 	&	43.76 	&	-1.151 	\\
2FGLJ1058.4+0133	&	1055+018	&	bzb	&	0.888 	&	2.642	&	2	&	8.41 	&	-2.19 	&	47.11 	&	47.42 	&	45.40 	&	44.51 	&	-0.011 	\\
2FGLJ1104.4+3812	&	Mkn 421	&	bzb	&	0.030 	&	0.2	&	6	&	8.15 	&	-1.80 	&	44.72 	&	44.88 	&	45.11 	&	41.70 	&	-2.567 	\\
2FGLJ1130.3-1448	&	1127?145	&	bzq	&	1.184 	&	1.89	&	1	&	9.18 	&	-2.76 	&	47.36 	&	47.68 	&	47.85 	&	45.77 	&	0.476 	\\
2FGLJ1136.7+7009	&	Mrk 180	&	bzb	&	0.045 	&	0.0827	&	6	&	8.21 	&	-1.45 	&	44.46 	&	43.85 	&	44.56 	&		&		\\
2FGLJ1159.5+2914	&	4C+29.45	&	bzq	&	0.724 	&	2.24	&	1	&	8.50 	&	-3.10 	&	46.33 	&	47.31 	&	44.25 	&	44.68 	&	0.068 	\\
2FGLJ1203.2+6030	&	GB6 J1203+6031	&	bzb	&	0.066 	&	0.146	&	2	&	8.05 	&	-1.48 	&	43.95 	&	43.95 	&	42.77 	&		&		\\
2FGLJ1217.8+3006	&	1215+303	&	bzb	&	0.130 	&	0.276	&	6	&	8.12 	&	-1.90 	&	45.51 	&	45.18 	&	45.67 	&		&		\\
2FGLJ1219.7+0201	&	PKS 1217+02	&	bzq	&	0.241 	&	0.257	&	2	&	8.87 	&	-1.89 	&	45.46 	&	45.14 	&	45.63 	&	44.83 	&	-0.154 	\\
2FGLJ1221.3+3010	&	1ES 1218+304	&	bzb	&	0.184 	&	0.0398	&	6	&	8.60 	&	-2.05 	&	45.39 	&	45.61 	&	45.81 	&		&		\\
2FGLJ1221.4+2814	&	1219+285	&	bzb	&	0.102 	&	0.94	&	2	&	7.70 	&	-1.91 	&	44.68 	&	45.18 	&	45.30 	&	42.25 	&	-1.564 	\\
2FGLJ1222.4+0413	&	4C+04.42	&	bzq	&	0.965 	&	0.665	&	2	&	8.31 	&	-2.65 	&	46.57 	&	47.36 	&	47.42 	&	44.91 	&	0.495 	\\
2FGLJ1224.9+2122	&	4C+21.35	&	bzq	&	0.432 	&	0.35	&	1	&	8.74 	&	-3.61 	&	46.06 	&	47.50 	&	43.91 	&	45.18 	&	0.333 	\\
2FGLJ1229.1+0202	&	1226+023	&	bzq	&	0.158 	&	26.4	&	2	&	8.51 	&	-2.58 	&	46.11 	&	46.34 	&	43.96 	&	45.54 	&	0.911 	\\
2FGLJ1258.2+3231	&	B2 1255+32	&	bzq	&	0.806 	&	0.894	&	2	&	8.50 	&	-2.31 	&	46.08 	&	46.46 	&	44.30 	&	44.41 	&	-0.197 	\\
2FGLJ1310.6+3222	&	1308+326	&	bzq	&	0.996 	&	1.97	&	2	&	8.48 	&	-2.99 	&	46.88 	&	47.56 	&	44.65 	&	44.98 	&	0.386 	\\
2FGLJ1317.9+3426	&	B2 1315+34A	&	bzq	&	1.056 	&	0.35	&	2	&	9.22 	&	-2.37 	&	46.35 	&	46.53 	&	46.75 	&	45.08 	&	-0.248 	\\
2FGLJ1326.8+2210	&	B2 1324+22	&	bzq	&	1.400 	&	1.134	&	3	&	9.25 	&	-2.77 	&	46.79 	&	47.69 	&	47.74 	&	44.93 	&	-0.429 	\\
2FGLJ1332.7+4725	&	B3 1330+476	&	bzq	&	0.668 	&	0.312	&	2	&	8.27 	&	-2.16 	&	45.91 	&	46.07 	&	44.14 	&	44.32 	&	-0.065 	\\
2FGLJ1419.4+3820	&	B3 1417+385	&	bzq	&	1.831 	&	0.514	&	3	&	8.62 	&	-2.73 	&	46.97 	&	47.58 	&	47.67 	&	45.10 	&	0.371 	\\
2FGLJ1428.0-4206	&	1424-418	&	bzq	&	1.522 	&	1.37	&	1	&		&	-3.02 	&	46.86 	&	48.43 	&	48.44 	&	44.95 	&		\\
2FGLJ1428.6+4240	&	1426+428	&	bzb	&	0.129 	&	0.0191	&	6	&	9.13 	&	-1.67 	&	45.15 	&	44.81 	&	43.64 	&		&		\\
2FGLJ1442.7+1159	&	1440+122	&	bzb	&	0.163 	&	0.0172	&	6	&	8.44 	&	-1.65 	&	45.10 	&	44.75 	&	43.62 	&		&		\\
2FGLJ1504.3+1029	&	PKS 1502+106	&	bzq	&	1.839 	&	0.56	&	1	&	9.11 	&	-2.66 	&	47.40 	&	49.12 	&	46.47 	&	45.23 	&	0.014 	\\
2FGLJ1512.2+0201	&	PKS 1509+022	&	bzq	&	0.219 	&	0.13	&	2	&	8.42 	&	-2.01 	&	45.11 	&	45.48 	&	45.63 	&	43.02 	&	-1.505 	\\
2FGLJ1517.7-2421	&	1514-241	&	bzb	&	0.049 	&	1.948	&	6	&	7.65 	&	-1.67 	&	44.29 	&	44.50 	&	44.71 	&		&		\\
2FGLJ1522.7-2731	&	PKS1519-273	&	bzb	&	1.294 	&	1.6	&	1	&	8.80 	&	-2.78 	&	47.16 	&	47.66 	&	45.00 	&	43.53 	&	-1.382 	\\
2FGLJ1535.4+3720	&	RGB J1534+372	&	bzb	&	0.142 	&	0.02	&	2	&	7.84 	&	-1.52 	&	44.01 	&	44.42 	&	43.05 	&		&		\\
2FGLJ1549.5+0237	&	1546+027	&	bzq	&	0.414 	&	1.147	&	2	&	8.62 	&	-2.27 	&	45.62 	&	46.26 	&	46.35 	&	44.80 	&	0.073 	\\
2FGLJ1550.7+0526	&	4C+05.64	&	bzq	&	1.422 	&	0.58	&	1	&	9.18 	&	-2.64 	&	47.13 	&	47.33 	&	47.54 	&	45.07 	&	-0.225 	\\
2FGLJ1608.5+1029	&	4C+10.45	&	bzq	&	1.226 	&	0.42	&	1	&	8.97 	&	-2.85 	&	46.80 	&	47.56 	&	44.78 	&	45.04 	&	-0.045 	\\
2FGLJ1613.4+3409	&	1611+343	&	bzq	&	1.397 	&	3.366	&	2	&	9.34 	&	-2.11 	&	47.40 	&	46.98 	&	45.43 	&	45.61 	&	0.154 	\\
2FGLJ1625.7-2526	&	1622-253	&	bzq	&	0.786 	&	1.16	&	1	&		&	-2.69 	&	46.55 	&	47.44 	&	44.80 	&	43.64 	&		\\
2FGLJ1626.1-2948	&	1622?297	&	bzq	&	0.815 	&	1.5	&	8	&	9.10 	&	-2.54 	&	47.02 	&	47.04 	&	44.80 	&		&		\\
2FGLJ1653.9+3945	&	Mkn 501	&	bzb	&	0.034 	&	0.491	&	6	&	8.84 	&	-1.66 	&	44.51 	&	44.46 	&	44.79 	&	42.20 	&	-2.746 	\\
2FGLJ1728.2+0429	&	1725+044	&	bzq	&	0.296 	&	0.98	&	2	&	7.90 	&	-2.11 	&	45.24 	&	45.78 	&	45.89 	&	44.07 	&	0.056 	\\
2FGLJ1728.2+5015	&	I Zw 187	&	bzb	&	0.055 	&	0.11	&	6	&	7.86 	&	-1.44 	&	44.48 	&	43.83 	&	44.57 	&		&		\\
2FGLJ1740.2+5212	&	1739+522	&	bzq	&	1.375 	&	1.9	&	2	&	9.32 	&	-2.81 	&	46.99 	&	47.83 	&	47.89 	&	45.16 	&	-0.274 	\\
2FGLJ1751.5+0938	&	4C+09.57	&	bzb	&	0.322 	&	1.695	&	2	&	8.41 	&	-2.10 	&	45.58 	&	46.26 	&	44.24 	&	43.70 	&	-0.828 	\\
2FGLJ1800.5+7829	&	CGRaBS J1800+7828	&	bzb	&	0.680 	&	1.436	&	4	&	8.26 	&	-2.25 	&	46.83 	&	47.08 	&	45.02 	&	44.85 	&	0.477 	\\
2FGLJ1806.7+6948	&	3C 371	&	bzb	&	0.051 	&	0.56	&	6	&	8.61 	&	-0.34 	&	44.40 	&	44.42 	&	44.37 	&	42.00 	&	-2.719 	\\
2FGLJ1824.0+5650	&	1823+568	&	bzb	&	0.664 	&	1.12	&	2	&	9.26 	&	-2.32 	&	46.59 	&	46.93 	&	44.77 	&	43.30 	&	-2.074 	\\
2FGLJ1848.5+3216	&	B2 1846+32	&	bzq	&	0.798 	&	0.508	&	2	&	8.04 	&	-2.46 	&	45.93 	&	46.85 	&	44.44 	&	44.58 	&	0.423 	\\
2FGLJ1849.4+6706	&	1849+670	&	bzq	&	0.657 	&	0.666	&	2	&	9.14 	&	-2.61 	&	45.98 	&	47.25 	&	47.27 	&	44.42 	&	-0.836 	\\
2FGLJ2000.0+6509	&	1ES 1959+650	&	bzb	&	0.047 	&	0.1817	&	6	&	8.10 	&	-1.55 	&	44.53 	&	44.52 	&	43.27 	&		&		\\
2FGLJ2009.5-4850	&	PKS 2005-489	&	bzb	&	0.071 	&	0.64	&	1	&	8.54 	&	-1.64 	&	44.92 	&	44.74 	&	43.50 	&	42.04 	&	-2.616 	\\
2FGLJ2143.5+1743	&	OX 169	&	bzq	&	0.211 	&	0.386	&	2	&	8.48 	&	-2.19 	&	45.06 	&	46.01 	&	46.06 	&	44.26 	&	-0.334 	\\
2FGLJ2148.2+0659	&	2145+067	&	bzq	&	0.990 	&	5.25	&	1	&	8.87 	&	-2.10 	&	47.00 	&	46.87 	&	45.14 	&	45.77 	&	0.782 	\\
2FGLJ2157.9-1501	&	2155-152	&	bzq	&	0.672 	&	2.654	&	3	&	7.59 	&	-2.34 	&	46.29 	&	46.44 	&	46.67 	&	43.68 	&	-0.022 	\\
2FGLJ2202.8+4216	&	BL Lac	&	bzb	&	0.069 	&	1.148	&	6	&	8.21 	&	-1.77 	&	45.14 	&	45.16 	&	43.68 	&	42.52 	&	-1.805 	\\
2FGLJ2204.6+0442	&	2201+044	&	bzb	&	0.027 	&	0.1313	&	6	&	8.10 	&	-1.20 	&	43.32 	&	43.11 	&	43.53 	&		&		\\
2FGLJ2211.9+2355	&	PKS 2209+236	&	bzq	&	1.125 	&	0.428	&	3	&	8.46 	&	-2.47 	&	46.14 	&	46.86 	&	44.47 	&	44.79 	&	0.212 	\\
2FGLJ2225.6-0454	&	2223-052	&	bzq	&	1.404 	&	6.935	&	3	&	8.42 	&	-2.64 	&	47.53 	&	47.74 	&	47.95 	&	45.60 	&	1.071 	\\
2FGLJ2229.7-0832	&	PKS 2227-08	&	bzq	&	1.560 	&	0.924	&	3	&	8.79 	&	-2.30 	&	47.20 	&	48.26 	&	46.00 	&	45.56 	&	0.657 	\\
2FGLJ2232.4+1143	&	2230+114	&	bzq	&	1.037 	&	1.44	&	1	&	8.78 	&	-2.68 	&	47.21 	&	47.56 	&	45.04 	&	45.75 	&	0.851 	\\
2FGLJ2236.4+2828	&	2234+282	&	bzb	&	0.795 	&	1.87	&	1	&	8.35 	&	-2.58 	&	46.47 	&	47.15 	&	44.65 	&	44.44 	&	-0.028 	\\
2FGLJ2253.9+1609	&	2251+158	&	bzq	&	0.859 	&	12.19	&	2	&	8.83 	&	-2.88 	&	47.22 	&	48.79 	&	45.92 	&	45.59 	&	0.646 	\\
2FGLJ2258.0-2759	&	2255-282	&	bzq	&	0.926 	&	2.58	&	1	&	9.04 	&	-2.65 	&	46.48 	&	47.35 	&	47.40 	&	45.84 	&	0.686 	\\
2FGLJ2327.5+0940	&	PKS 2325+093	&	bzq	&	1.841 	&	0.315	&	2	&	9.03 	&	-2.97 	&	47.06 	&	48.14 	&	45.21 	&	45.20 	&	0.061 	\\
2FGLJ2334.3+0734	&	TXS 2331+073	&	bzq	&	0.401 	&	0.605	&	3	&	8.37 	&	-2.14 	&	45.44 	&	45.87 	&	46.01 	&	44.93 	&	0.448 	\\
2FGLJ2347.9-1629	&	2345-16	&	bzq	&	0.576 	&	1.951	&	3	&	8.60 	&	-2.36 	&	46.17 	&	46.50 	&	46.67 	&	44.36 	&	-0.350 	\\

\enddata
\tablecomments{(1)  Dodson et al. (2008); (2) Laurent-Muehleisen et al. (1997); (3) Cooper et al. (2007); (4) Gu et al. (2009); (5) Caccianiga et al. (2004); (6) Giroletti et al. (2004); (7) Jorstad et al. (2004).bzb=BL Lacs, bzq=FSRQs. }
\end{deluxetable}

\end{document}